\documentclass[twocolumn,aps,prl,amsmath,superscriptaddress,notitlepage,longbibliography]{revtex4-1}
\usepackage{amssymb}
\usepackage{mathrsfs}
\usepackage{graphicx}
\usepackage{float}
\usepackage[caption=false]{subfig}
\usepackage{epstopdf}

\usepackage[normalem]{ulem}
\usepackage{verbatim}
\usepackage{xcolor}
\usepackage{bm}

\usepackage[utf8x]{inputenc}

\def\CH{\textcolor{black}}
\def\blue{\textcolor{blue}}
\def\red{\textcolor{red}}

\begin{document}

\def\qv{\vec{q}}
\def\red{\textcolor{red}}
\def\blue{\textcolor{blue}}
\def\magenta{\textcolor{magenta}}
\def\apricot{\textcolor{Apricot}}

\def\GJ{\textcolor{black}}
\def\TT{\textcolor{ForestGreen}}
\definecolor{ora}{rgb}{1,0.45,0.2}
\def\LH{\textcolor{black}}

\newcommand{\norm}[1]{\left\lVert#1\right\rVert}
\newcommand{\ad}[1]{\text{ad}_{S_{#1}(t)}}

\title{Critical non-Hermitian Skin Effect}

\author{Linhu Li}
\email{phylli@nus.edu.sg}
\affiliation{Department of Physics, National University of Singapore, Singapore 117542}
\author{Ching Hua Lee}
\email{phylch@nus.edu.sg}
\affiliation{Department of Physics, National University of Singapore, Singapore 117542}
\affiliation{Institute of High Performance Computing, A*STAR, Singapore 138632}
\author{Sen Mu}
\email{senmu@u.nus.edu}
\affiliation{Department of Physics, National University of Singapore, Singapore 117542}
\author{Jiangbin Gong}
\email{phygj@nus.edu.sg}
\affiliation{Department of Physics, National University of Singapore, Singapore 117542}

\date{\today}

\date{\today}
\begin{abstract}

\GJ{This work uncovers a \CH{new} class of criticality where eigenenergies and eigenstates of non-Hermitian lattice systems jump discontinuously across a critical point in the thermodynamic limit, \CH{unlike established Hermitian and non-Hermitian critical scenarios where spectrum  remains continuous across a transition.}
 Such critical behavior, dubbed the ``critical skin effect",  is rather generic, 
\CH{occuring whenever subsystems with dissimilar non-Hermitian skin localization lengths are coupled, however weakly. }
Due to the existence of this criticality, the thermodynamical limit and the zero-coupling limit cannot be exchanged, thus challenging the celebrated generalized Brillouin zone (GBZ) approach when applied to finite-size systems. As manifestations of the critical skin effect in finite-size systems, we present stimulating examples with anomalous scaling behavior regarding spectrum, correlation functions, entanglement entropy,} \CH{and scale-free wavefunctions that decay exponentially rather than power-law. More spectacularly, topological in-gap modes can even be induced by changing the system size. } 
\end{abstract}

\maketitle

\noindent{\textit{Introduction.--}}
\GJ{Lying at the boundary between distinct phases}, critical systems exhibit a wide range of interesting universal properties from divergent susceptibilities to anomalous scaling behavior. 
They have broad ramifications in conformal and statistical field theory~\cite{coniglio1980clusters,hu1984percolation,aizenman1987phase,boulatov1987ising,zamolodchikov1987conformal,zamolodchikov1989exact,cardy1992critical,oshikawa1996boundary,dziarmaga2005dynamics}, Schramm-Loewner evolution~\cite{gruzberg2006stochastic,kozdron2009using,stevenson2011domain,werner2007lectures}, entanglement entropy (EE)~\cite{vidal2003entanglement,korepin2004universality,larsson2005entanglement,ryu2006aspects,barthel2006entanglement,laflorencie2006boundary,swingle2010entanglement,swingle2013universal,lee2015exact,swingle2016area, chang2019entanglement} and many other contexts.  Recently, concepts \GJ{crucial} to criticality - like band gaps and \GJ{localization} -  have been challenged by studies of non-Hermitian systems~\cite{Bender1998nonH,bender2007making,NHbook,gong2018topological,kawabata2019symmetry} \CH{exhibiting
exceptional points~\cite{berry2004EP,dembowski2004encircling,Rotter2009non,jin2009solutions,longhi2010pt,heiss2001chirality,heiss2012EP,xu2016topological,Hassan2017EP,Hu2017EP,shen2018topological,wang2019arbitrary,ghatak2019new,miri2019exceptional,zhang2020non,yuce2020non,jin2019hybrid,kawabata2019classification} or the non-Hermitian skin effect (NHSE), which are characterized by enigmatic bulk-boundary correspondence (BBC) violations, robust directed amplifications, discontinuous Berry curvature and anomalous transport behavior~\cite{Lee2016nonH,xiong2018does,kunst2018biorthogonal,yao2018edge,yokomizo2019non,lee2018tidal,Lee2019anatomy,song2019non,song2019realspace,li2019geometric,borgnia2020nonH,zhang2019correspondence,yoshida2019mirror,yang2019auxiliary,lee2019unraveling,longhi2019probing,luo2020skin}.}

\begin{figure}
\includegraphics[width=\linewidth]{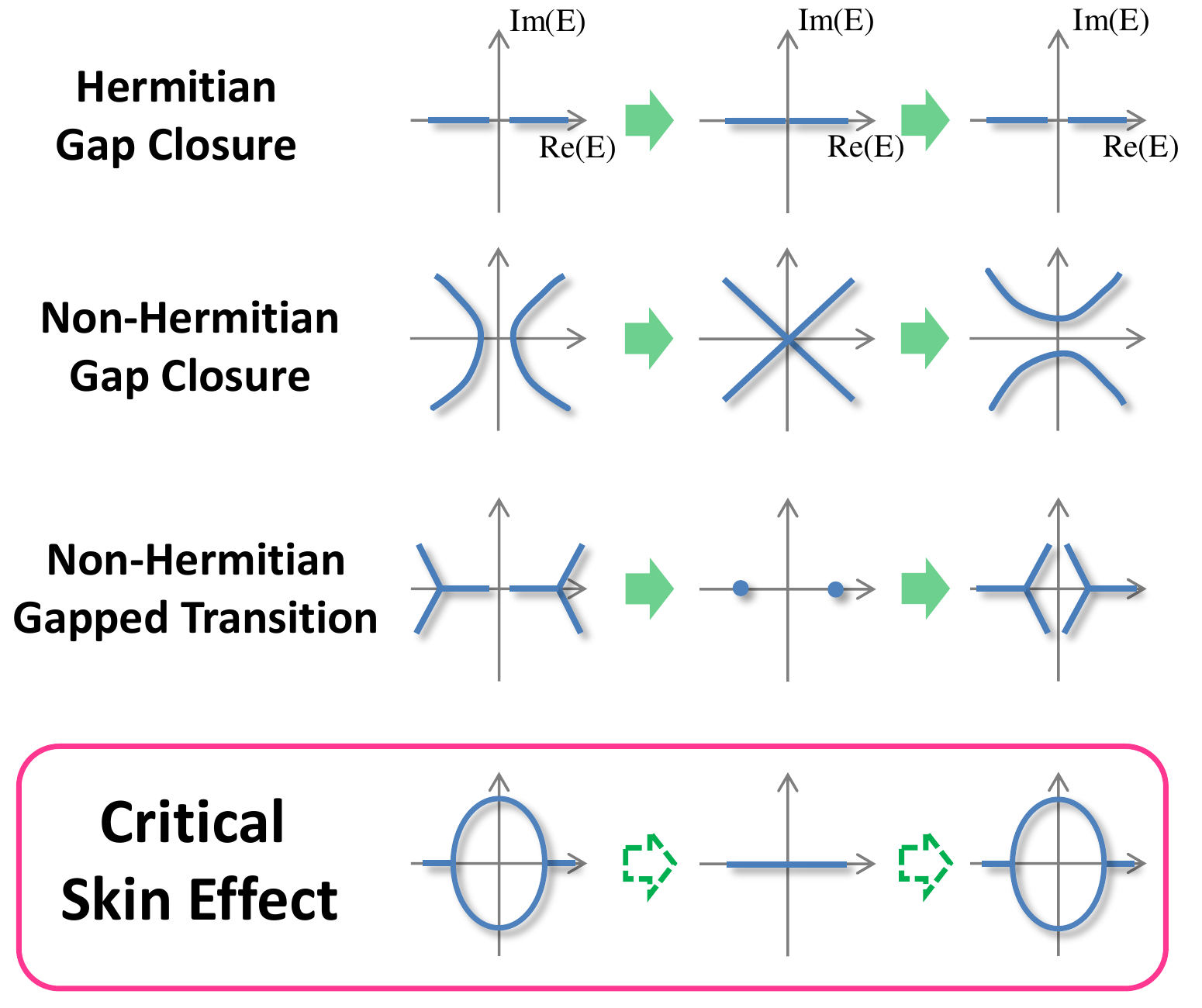}
\caption{Four different types of critical transitions. Hermitian phase transitions (Top) are marked by gap closures along the real line. In non-Hermitian cases (2nd to 4th rows, axis labels omitted), spectral phase transitions can take more sophisticated possibilities in the 2D complex energy plane. For instance, the spectral topology can change under line gap closures (2nd Row) or shrink to a point and re-emerge in a different topological configuration (3rd row), without the gap ever closing~\cite{lee2019unraveling}.  The spectrum continuously passes through a gapless or point-like regime in the first three cases.   The CSE (bottom row), however, is special in that  OBC spectrum the thermodynamic limit, denoted $E_\infty$, jumps discontinuously from \GJ{one configuration (left), to a different one (middle), and to another (right) as certain parameter changes from $-\epsilon$ to $0$ (critical border), and to $\epsilon$, for an arbitrarily small $\epsilon$}.  
}
\label{fig:comparison}
\end{figure}

\GJ{We uncover here a \CH{new} class of criticality, dubbed the ``critical skin effect'' (CSE), where the eigenenergies and eigenstates  in the thermodynamic limit ``jump'' discontinuously across the critical point.} \CH{This is distinct from previously known phase transitions (Hermitian and non-Hermitian) [Fig.~\ref{fig:comparison}]}, where the eigenenergy spectrum can be continuously interpolated across the two bordering phases. A CSE transition, by contrast, is characterized by a discontinuous jump between two different complex spectra along with two different sets of eigenstates. \CH{As elaborated below, this behavior appears generically whenever systems of dissimilar NHSE localization lengths are coupled, no matter how weakly~\footnote{Not all subsystems need to be non-Hermitian; indeed, the CSE even occurs if a Hermitian chain is coupled a chain with NHSE}.} 
Importantly, at experimentally accessible finite system sizes~\footnote{Non-Hermitian system, particularly those exhibiting the NHSE, exhibit divergences in local density of states in the thermodynamic limit.}, the jump smooths out into an interpolation between the two phases in a strongly size-dependent manner, such that the system  may exhibit qualitatively different properties i.e. real vs. complex spectrum or presence/absence of topological modes at different system sizes. Being strongly affected by minute perturbations around the critical point, such behavior may prove useful in sensing applications~\cite{brandenbourger2019non,schomerus2020nonreciprocal}.


\noindent{\textit{CSE as a limitation of the GBZ.--}}
\GJ{In non-Hermitian systems with unbalanced gain and loss, spectra under periodic boundary conditions (PBCs) and open boundary conditions (OBCs) can be very different~\cite{Lee2016nonH,xiong2018does,yao2018edge,PhysRevLett.124.066602,Lee2019anatomy}. 
Indeed, under OBC,  eigenstates due to NHSE can exponentially localize at a boundary, in contrast to Bloch states under PBCs. 
This also explains the possible violation of the BBC, taken for granted in Hermitian settings.}

The celebrated GBZ formalism aims to restore the BBC through a complex momentum deformation~\cite{yao2018edge,yokomizo2019non,Lee2019anatomy,yang2019auxiliary,PhysRevLett.124.086801,lee2019unraveling}. Rigorously applicable for 
bounded but infinitely large systems, it has however been an open question whether the GBZ can still accurately describe finite-size systems. The GBZ of a momentum-space Hamiltonian $H(z)$, $z=e^{ik}$ can be derived from its characteristic Laurent polynomial (energy eigenequation)
\begin{eqnarray}
f(z,E):=\det [H(z)-E]=0\label{eigenequation},
\end{eqnarray}
where $E$ is the eigenenergy. While the ordinary BZ is given by the span of allowed real quasimomenta $k$, the GBZ is defined by the complex analytically-continued momentum $k\rightarrow k+i\kappa(k)$, with the NHSE inverse decay length $\kappa(k)=-\log|z|$ determined by the smallest complex deformation $z\rightarrow e^{ik}e^{-\kappa(k)}$ such that $f(z,E)$ possesses a pair of zeros $z_\mu$, $z_\nu$ satisfying $|z_\mu|=|z_\nu|$ for the same $E$~\cite{yao2018edge,Lee2019anatomy,lee2019unraveling}. Due to the double degeneracy of states with equal asymptotic decay rate at these $E$, there exist a pair 
of eigenstates $\psi_\mu,\psi_\nu$ that can superpose to satisfy OBCs i.e. zero net amplitude at both boundaries. 
As such, provided that the characteristic polynomial \CH{cannot be made reducible by adding a small perturbation}, the OBC spectrum in the thermodynamic limit (denoted as $E_\infty$) can be obtained from the PBC spectrum via $E(e^{ik})\rightarrow E(e^{ik}e^{-\kappa(k)})$, apart from isolated topological modes. Thus it is often claimed that the BBC is ``restored'' in the GBZ defined by $k\rightarrow k+i\kappa(k)$ or, at the operator level, with the surrogate Hamiltonian $H(e^{ik})\rightarrow H(e^{ik}e^{-\kappa(k)})$~\cite{lee2019unraveling}.  In general, different $E$ (energy band) solutions can admit different functional forms of $\kappa(k)$, leading to band-dependent GBZs that have recently also been described with the auxiliary GBZ formalism~\cite{yang2019auxiliary}.
\CH{Since $e^{ik}e^{-\kappa(k)}$ is generically non-analytic, it represents effectively non-local hopping terms~\cite{lee2019unraveling}. As such, the GBZ description challenges the very notion of locality, which is central to critical systems, by effectively ``unraveling'' the real-space eigenstate accumulation through replacing local hoppings with effectively non-local ones. }


Due to the robustness of the  NHSE, eigenspectra predicted from the GBZ typically converge rapidly to the exact numerically obtained OBC spectra even for small system sizes($\mathcal{O}(10^1)$ sites)~\footnote{In principle, the convergence should be exact in the thermodynamic limit. But in practical computations, floating point errors $\epsilon$ are continuously amplified as they propagate across the system, and we expect accurate numerical spectra only when $L<-\log(\epsilon)/\text{max}(\kappa) $. }. However, this numerical agreement fails spectacularly near a critical point where $f(z,E)$ changes from being reducible to irreducible. To understand the significance of this algebraic property of reducibility, consider a set of coupled irreducible subsystems described by the characteristic polynomial
\begin{equation}
f(z,E)=f_0+\prod_i f_i(z,E),
\label{fzE}
\end{equation}
where $f_i(z,E)$ is the characteristic polynomial of the $i$-th subsystem, and $f_0$ is a constant that represents the simplest possible form for the subsystem coupling. When $f_0=0$, $f(z,E)$ completely factorizes into irreducible polynomials, as expected from a Hamiltonian $H(z)$ that block-diagonalizes into irreducible sectors associated with the individual $f_i(z,E)$'s. In particular, the OBC spectrum of this completely decoupled scenario is derived from the independent $\kappa_i(k)$'s of each subsystem, each determined by $z_\mu,z_\nu$ from the \emph{same} subsystem.

Yet, a nonzero coupling $f_0$, no matter how small, can have dramatic physical consequences. \CH{For arbitrarily small $f_0\ne 0$}, the different sectors can hybridize significantly if the $f_i$'s are different~\footnote{If two $f_i$'s are equivalent, $f_0+f_i^2=(f_i+i\sqrt{f_0})(f_i-i\sqrt{f_0})$ is still reducible.}. Indeed, such hybridization is \emph{inevitable} in the thermodynamic limit, with OBC eigenstates formed from superpositions of eigenstates $\psi_\mu,\psi_\nu$ from dissimilar subsystems, each corresponding non-Bloch momenta $-i\log z_{\mu/\nu}$. Hence the GBZs i.e. $\kappa(k)$'s of the coupled system, which are defined in the thermodynamic limit, are thus determined by all pairs of $|z_\mu|=|z_\nu|$ not necessarily from the same subsystem.  \GJ{Therefore, the GBZs in the coupled case, no matter how small is $f_0$}, can differ from the decoupled GBZs at $f_0=0$. \GJ{That is,  the thermodynamic limit and the $f_0\rightarrow 0$ limit are \CH{\emph{not}} exchangeable.}
 However, since an actual finite physical system cannot possibly possess very different spectrum and band structure upon an arbitrarily small variation in its system parameter, the GBZ picture must  \CH{be inapplicable in describing} \GJ{such finite-size systems} in the presence of CSE.

\begin{figure}
\includegraphics[width=\linewidth]{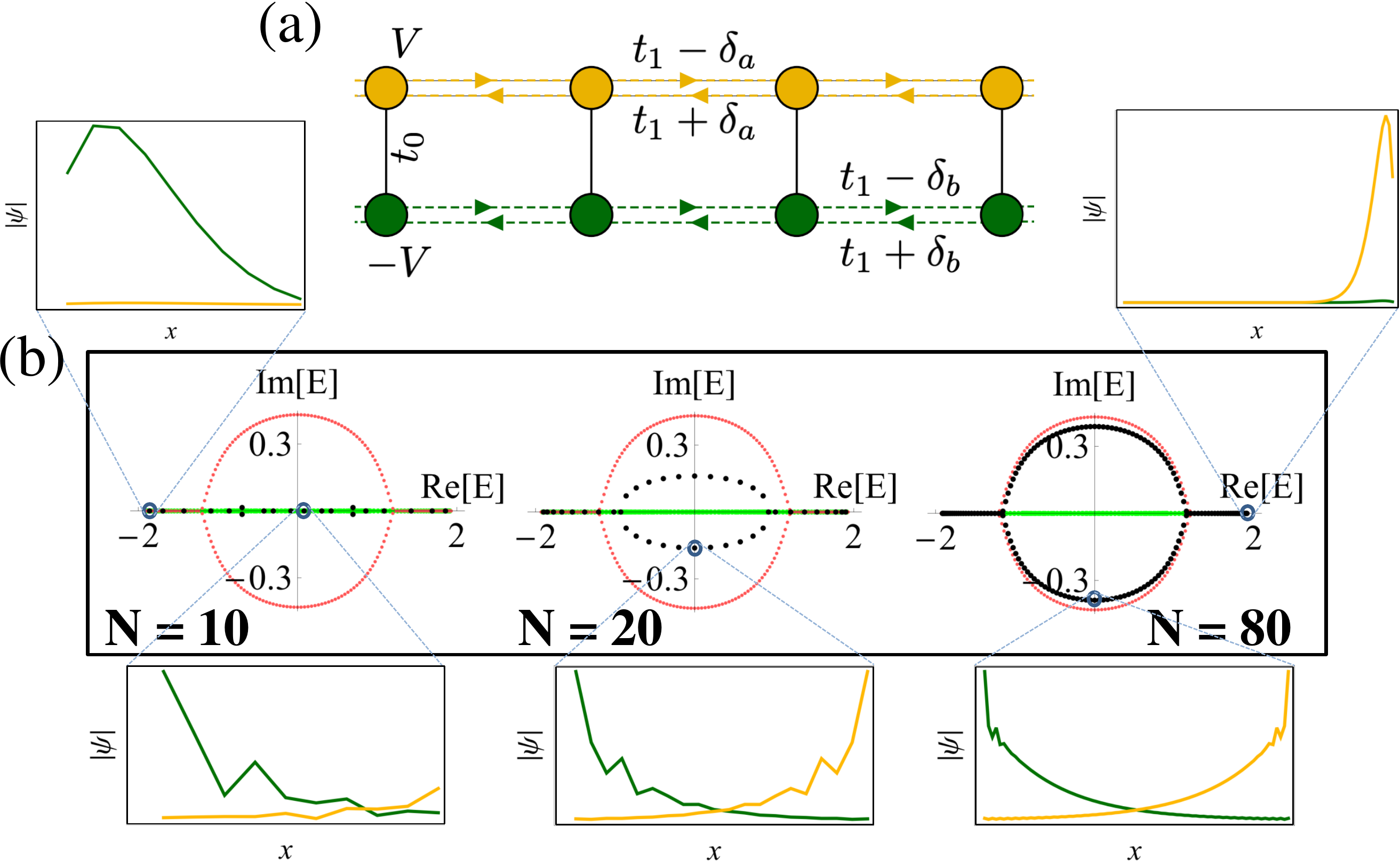}
\caption{
(a) The two chain model [Eq.~\ref{two-chain_k}] with hopping asymmetry in chains $a,b$ denoted by $\delta_{a/b}$, and on-site energy offset $\pm V$. A small inter-chain $t_0$ can cause significant coupling when $\delta_a\neq \delta_b$.
(b) OBC spectra (black dots) and eigenstate profiles (insets) at $N=10,20$ and $80$ unit cells and coupling $t_0=0.01$, showing very different spectral behavior at different system sizes $N$. At small $N\approx 10$, coupling effects are negligible, with the spectrum coinciding with the real OBC $E_\infty$ spectrum (green) in the decoupled thermodynamic limit. As $N$ increases, the spectrum gradually approaches the OBC $E_\infty$ spectrum (red) for the coupled thermodynamic limit, with hybridization becoming sharper. Parameters are $t_1=0.75$, $\delta_a=-\delta_b=0.25$ and $V=0.5$.}
\label{fig:two-chain}
\end{figure}

\noindent{\textit{Anomalous finite-size scaling from CSE.--}}
For illustration, we turn to a minimal example of two coupled non-Hermitian 1D Hatano-Nelson chains~\cite{HN1996prl,HN1997prb,HN1998prb} each containing only non-reciprocal (unbalanced) nearest neighbor (NN) hoppings [Fig.~\ref{fig:two-chain}(a)]. Its Hamiltonian reads
\begin{eqnarray}
H_\text{2-chain}(z)=\left(\begin{matrix}
g_a(z) & t_0\\
t_0 & g_b(z)\\
\end{matrix}\right)\label{two-chain_k}
\end{eqnarray}
with $g_a(z)=t_a^+z+t_a^-/z+V$ and $g_b(z)=t_b^+z+t_b^-/z-V$, $t_{a/b}^\pm=t_1\pm\delta_{a/b}$ being the forward/backward hopping of chains $a$ and $b$.  This model can be also realized with a reciprocal system with skin effect in a certain parameter regime \cite{SuppMat}. When $t_0=0$, the two chains are decoupled, and the characteristic polynomial is reducible as $f(z,E)=[g_a(z)-E][g_b(z)-E]$. Each factor $f_{a/b}(z,E)=g_{a/b}(z)-E$ determines the skin eigensolutions of its respective chain. However, even an infinitesimal coupling $t_0\ne 0$ generically makes $f(z,E)$ irreducible. \GJ{Specifically, consider} the simple case of 
$t^+_a=t^-_b=1$ and $t_a^-=t_b^+=0$. \GJ{Without couplings ($t_0=0$), the two chains under OBC respectively yields a  Jordan-block Hamiltonian matrix in real space, with the spectrum given by $E=\pm V$. Because the eigenstates of the decoupled chains are exclusively localized at the first or the last site, their GBZs collapse~\cite{PhysRevLett.124.066602}}. By contrast, for any $t_0\ne 0$, $f(z,E)=E^2-E(z+z^{-1})+(z+V)(z^{-1}-V)-t_0^2$
is irreducible (here $-t_0^2=f_0$ from Eq.~\ref{fzE}), \GJ{insofar as} the eigenenergy roots 
$E=\cos k\pm \sqrt{t_0^2+(V+i\sin k)^2}$ are no longer Laurent polynomials in $z=e^{ik}$ that can be separately interpreted as de facto subsystems with local hoppings~\footnote{In higher degree polynomials, an algebriac expression for $z$ may not even exist as implied by the Abel-Ruffini theorem.}.  Importantly, the corresponding OBC $E_\infty$ spectrum and the GBZ for $t_0\ne 0$ are now qualitatively different. As derived in the Supplementary Material~\cite{SuppMat}, setting $|z_a|=|z_b|$ gives OBC spectrum (in the thermodynamic limit): $E_\infty^2=\frac{1-\eta^2}{1+\eta^2}+V^2+t_0^2\pm 2\sqrt{t_0^2-\eta^2+\eta^2t_0^2}/(1+\eta^2)$, with $\eta\in \mathbb{R}$. \GJ{Clearly, even one now takes the $t_0\rightarrow 0$ limit, $E_\infty^2$ only simplifies to $E_\infty^2\rightarrow V^2+\frac{1\pm i\eta}{1\mp i\eta}$, which is {\it not} the above-mentioned OBC spectrum of the two decoupled chains}.  \GJ{Likewise, the $t_0\rightarrow 0$ limit of the coupled GBZ, which can be shown to be the locus of $z=\pm\sqrt{V^2+e^{i\theta}}-V$, $\theta\in [0,2\pi]$, has nothing in common with the collapsed GBZs of the decoupled case.}


This paradoxical singular behavior is manifested as anomalous scaling behavior in finite-size systems \GJ{that are more relevant} to experimental setups. The discontinuous critical transition illustrated above becomes a smooth crossover between the different OBC $E_\infty$ solutions. As the size $N$ of a coupled system is varied, its physical OBC spectrum interpolates between the decoupled and coupled OBC $E_\infty$ solutions. As illustrated in Fig.~\ref{fig:two-chain}(b) for the 2-chain model Eq.~\ref{two-chain_k} at small coupling $t_0=0.01$ (with $t_1=0.75$ and $\delta_a=-\delta_b=0.25$ for well-defined skin modes), the OBC spectrum (black dots) changes dramatically from $N=10$ to $80$ unit cells. For small $N=10$, it approximates the OBC $E_\infty$ (green) for $t_0=0$ lying on the real line, while at large $N=80$, it converges towards the true OBC $E_\infty$ (red curve) with nonzero coupling. Indeed, the eigenstates for $N=10$ are almost entirely decoupled across the two chains, while those for $N=80$ are maximally coupled/decoupled depending on whether they approach the red/green $E_\infty$ curves. In the intermediate $N=20$ case, the OBC spectrum lies far between the two $E_\infty$'s, and cannot be characterized by their associated GBZs.

\begin{figure}
\subfloat[]{\includegraphics[width=.53\linewidth]{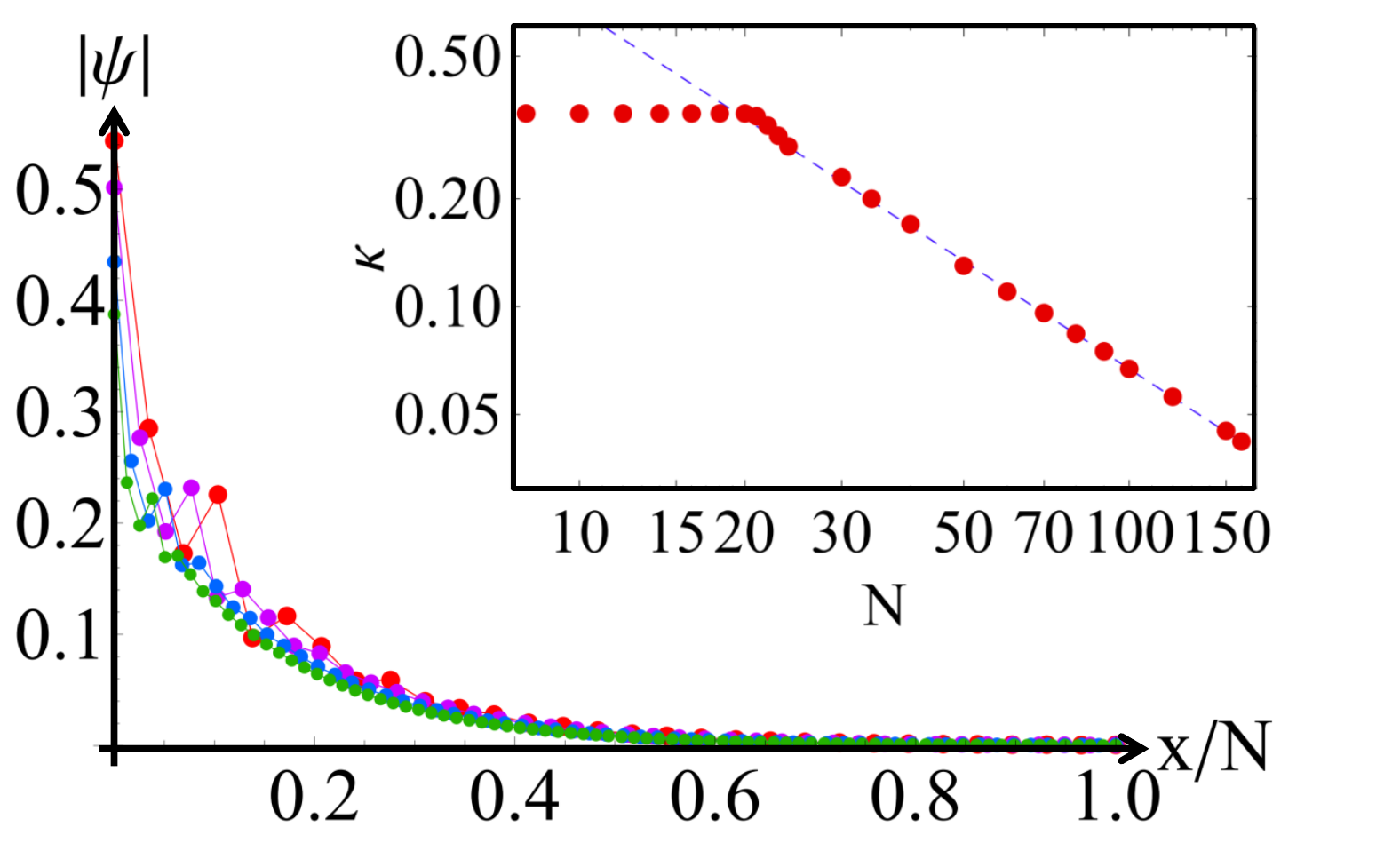}}
\subfloat[]{\includegraphics[width=.47\linewidth]{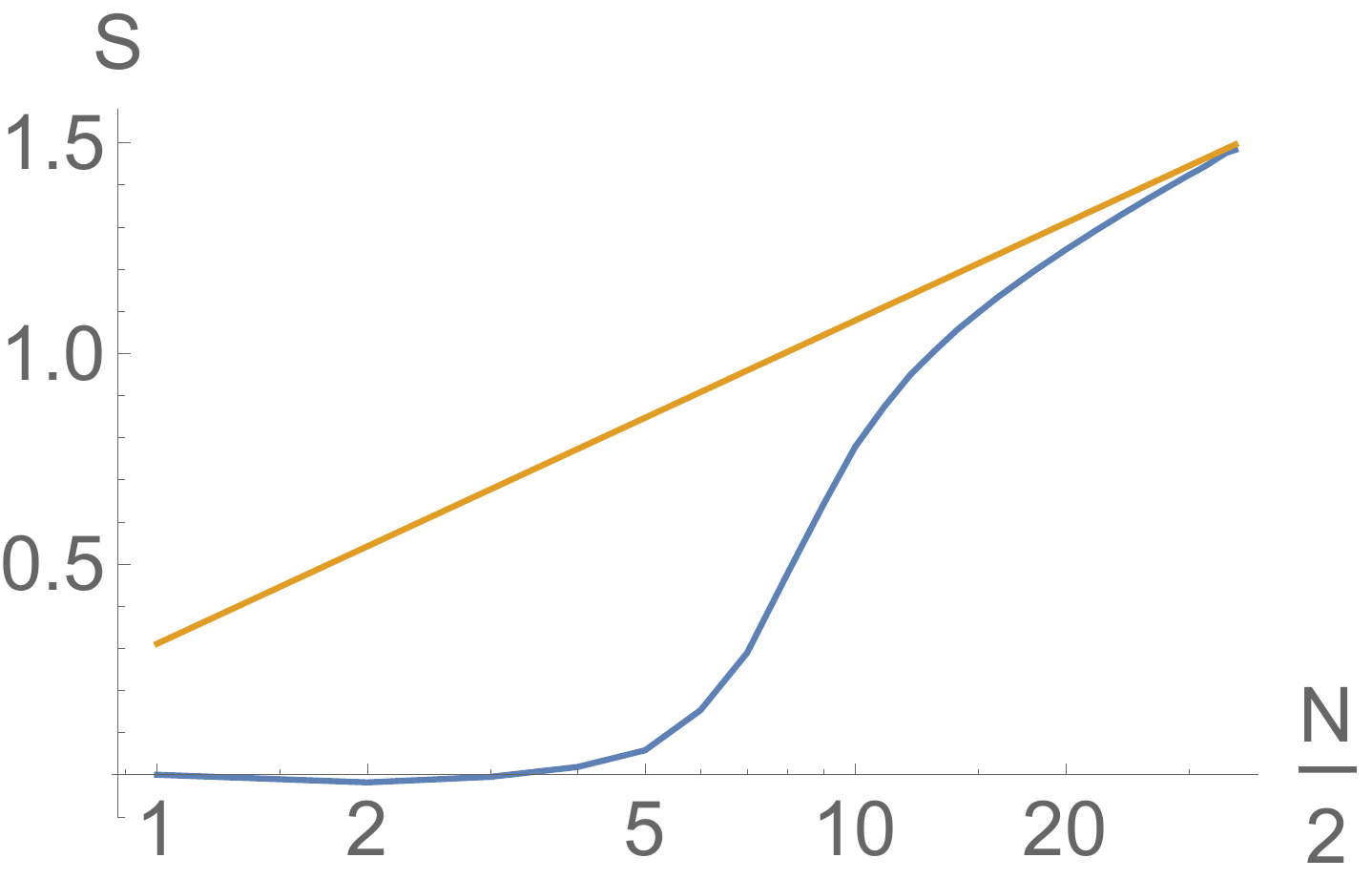}}
\caption{
(a) Scale-free OBC skin eigenstate of the largest $\text{Im}[E]$ eigenenergy of $H_\text{2-chain}$ at system sizes $N=20,40,60$ and $80$ (red,purple,blue,green). Its rescaled profile, despite decaying \emph{exponentially} rather than power-law, remains invariant across different $N$. This scale invariance persists in the $N>20$ regime, and is due to the $N^{-1}$ decay (dashed line) of the inverse skin depth (red dots), as plotted in the inset.  
Parameters follow Fig.~\ref{fig:two-chain}'s, except with $t_0=10^{-3}$.  (b) EE $S$ (blue) of a half-filled OBC $H_\text{2-chain}$ at odd system sizes $N$, with real-space cut at $\lfloor \frac{N}{2}\rfloor$ and parameters $t_1=0.58$, $V=1$, $t_0=0.4$ and $\delta_a=-\delta_b=0.25$. It saturates near zero in the gapless decoupled small $N$ regime, but scales like $\sim \frac1{3}\log N$ (yellow) in the gapless coupled large $N$ regime.}
\label{fig3}
\end{figure}

Let us now explain the above-observed dramatic size-dependent spectrum via the competition between dissimilarly accumulated skin modes and the couplings across them.
The general conditions for such are unveiled in Sec. I.a of~\cite{SuppMat}.
In our model [Eq.~\ref{two-chain_k}], the inverse \GJ{decay lengths} in chains $a,b$ are given by
$\kappa_{a/b}=\frac{1}{2}\log (t^+_{a/b}/t^-_{a/b})$,
which will be dissimilar as long as $\delta_a\neq \delta_b$. After performing a similarity transform that rescales each site $j$ by a factor of $e^{j\kappa_b}$, chain $b$ becomes reciprocal with $\kappa'_b=0$ while chain $a$ has a rescaled \GJ{inverse decay length} $\kappa'_a=\kappa_a-\kappa_b$. If $\kappa'_a\neq 0$, chain $a$ always possesses exponentially growing skin modes scaling like $~e^{\kappa'_aN}$ at one end. As such, the coupling $t_0$, even if extremely small, still affects the spectrum and eigenstates  \GJ{dramatically as the system size $N$ increases.}\red{
}

\noindent\textit{Scale-free exponential wavefunctions.--}
A hallmark of conventional critical systems is scale-free power-law behavior, particularly in the wavefunctions. Interestingly,  such scale-free behavior can also be found in the exponentially decaying wavefunctions i.e. skin modes. Shown in Fig.~\ref{fig3}(a) are the profiles of the slowest decaying eigenstates $\psi(x)$ of $H_\text{2-chain}$ at different system sizes $N=20,40,60$ and $80$, with the horizontal axis normalized by $N$.  \GJ{These featured eigenstates belong to the top of the central black ring in Fig.~\ref{fig:two-chain}(b), with their distance from the coupled OBC $E_\infty$ ring (red) decreasing as $\sim N^{-1}$.}
Unlike usual exponentially decaying wavefunctions with fixed spatial decay length, here $|\psi(x)|\sim e^{-\kappa x}$ with $\kappa\sim N^{-1}$ [Fig.~\ref{fig3}(b)], such that the overall profile $\psi(x)$ has no fixed length scale. Such unique scale-free eigenmodes result from the slow critical migration of the eigenstates between $E_\infty$ solutions [Fig.~\ref{fig:two-chain}(a) inset].

\vspace{0.1cm}
\noindent\textit{Anomalous correlations and entanglement entropy.--}
The CSE can also violate the usual logarithmic scaling of the EE~\cite{Cardy2004,Cardy2009,Gioev2006,Eisert2010area}, since the OBC spectrum can be gapped at some system sizes, and gapless at others. Consider for instance the OBC $H_\text{2-chain}$ [Eq.~\ref{two-chain_k}] with parameters chosen to gap out the OBC spectrum at small system sizes $N$~\cite{SuppMat}. With all $\text{Re}[E]<0$ states occupied by spinless free Fermions, the real-space entanglement entropy $S$ (blue curve in [Fig.~\ref{fig3}(c)]) exhibits a crossover from the decoupled gapped regime at $N\leq 5$ to the gapless regime $N>20$, where it approaches the usual $\frac1{3}\log N$ behavior (yellow line). In generic CSE scenarios with multiple competing OBC $E_\infty$ loci, $S$ can scale differently at different system size regimes, choices of fillings and entanglement cuts, challenging the notion of a single well-defined scaling behavior. As shown in the Supplementary Material~\cite{SuppMat}, The two-Fermion correlator $\langle \psi(1)\psi(x)\rangle$ characterizing the EE also crossovers from rapid exponential decay at small $N$ to $1/x$ power-law decay at large $N$. Remarkably, the probability of finding another Fermion nearby generally \emph{increases} drastically when the system is enlarged (with filling fraction maintained).

\begin{figure}
\includegraphics[width=1\linewidth]{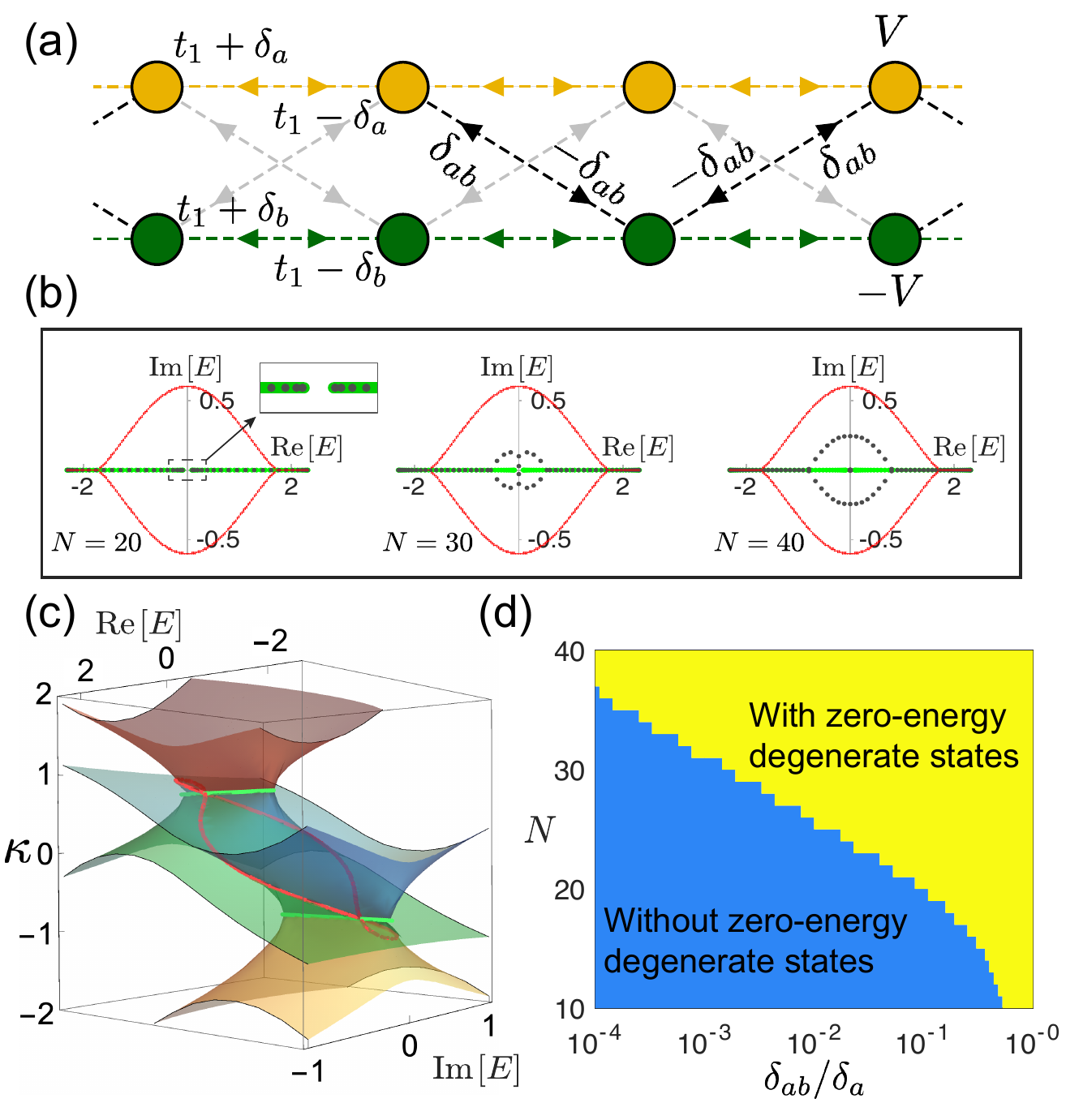}
\caption{
(a) Sketch of the $H_\text{CS-SSH}$ model with cross inter-chain non-reciprocal couplings $\pm\delta_{ab}$.
(b). OBC spectra (black dots) at $N = 20$, $30$ and $40$ unit cells and coupling $\delta_{ab}=0.5\times 10^{-3}$. The majority of the spectrum behaves similarly as the model in Fig.~\ref{fig:two-chain}(b), except for a pair of topological edge states emerge within the point gap at zero energy.
The OBC $E_\infty$ spectrum is given by green and red colors in the decoupled and coupled thermodynamic limit respectively.
Other parameters are $\delta_a=-\delta_b=0.5$, $t_1=0.75$, and $V=1.2$.
(c) $\kappa$ solutions (red, blue, green, and yellow surfaces) of $f(z,E)=0$ as a function of the complex energy, with the same parameters in (b). Intersecting regions (green and red dotted lines) give the OBC skin solutions of the system in the thermodynamic limit. Among them, green and blue lines correspond to the skin solutions of two decoupled chains at $\delta_{ab}=0$. The solutions of red curves emerge at a small but nonzero $\delta_{ab}$, and the skin solutions of the weakly coupled system is given by the intersecting regions with smallest $|\kappa|$, i.e. the red loop in the center and green lines at the two ends with large and small ${\rm Re}[E]$.
(d) \GJ{Emergence of in-gap degenerate modes} as a function of  $\delta_{ab}/\delta_{a}$ and $N$ with $\delta_a=-\delta_b=0.5$, $t_1=0.75$, $V=1.2$, with the plotted boundary scaling logarithmically with $N$.}
\label{fig:SSH}
\end{figure}

\noindent\textit{Size-dependent topological modes.--}
\GJ{Topological modes are usually associated with bulk invariants in the thermodynamic limit, with finite-size effects playing a diminishing role in the face of topological robustness}.  The CSE here can \GJ{cause topological edge modes to appear only at certain system size regimes}. Consider replacing the non-reciprocal intra-chain couplings of our $H_\text{2-chain}$ model with inter-chain couplings with non-reciprocity $\pm\delta_{ab}$ between adjacent unit cells [Fig.~\ref{fig:SSH}(a)], as described by the following CSE Su-Schrieffer-Heeger (SSH) model:
\begin{eqnarray}
H_\text{CSE-SSH}(z)&=&h_y(z)\sigma_y+h_z(z)\sigma_z+h_0(z)\mathbb{I}
\label{Hk_shift}
\end{eqnarray}
where $h_y(z)=i\delta_{ab}(z+1/z)$, $h_z(z)=V+\delta_-(z-1/z)$, and $h_0(z)=t_1(z+1/z)+\delta_+(z-1/z)$, with $\delta_\pm=(\delta_a\pm\delta_b)/2$. $H_\text{CSE-SSH}$ is so named because interestingly, at $\delta_-=\delta_{ab}$, it can be transformed via a basis rotation $\sigma_z\rightarrow \sigma_x$ into an extended (SSH) model~\cite{SSH} with non-reciprocal inter-cell couplings given by $\pm2\delta_-$ and a uniform next-nearest neighbor hopping given by $t_1\pm\delta_+$~\cite{SuppMat}, which is known to possess a topologically nontrivial phase.

When $\delta_{ab}=0$, the system is decoupled into two Hatano-Nelson chains which must be topologically trivial. \GJ{The OBC spectrum $E_\infty$ in the decoupled case} and the associated inverse decay length $\kappa$ are shown in Figs.~\ref{fig:SSH}(b) and (c)(green curves), with positive/negative $\kappa$ corresponding to skin modes accumulating population at opposite boundaries. Also shown in
in Figs.~\ref{fig:SSH}(b) and (c)(red curves) are $E_\infty$ in the coupled case and the corresponding $\kappa$ for the  hybridized skin modes. With small $N=20$ unit cells in Fig.~\ref{fig:SSH}(b), the finite-size OBC spectrum (gray dots) qualitatively agrees with the decoupled $E_\infty$ (green), with a \GJ{real-valued gap at $E=0$ along the $\text{Im}[E]=0$ axis (inset)}. Upon the size increase to $N=30$ and then to $N=40$, \GJ{such a gap first closes on the complex plane} and then develops into a point gap with two zero-energy degenerate modes lying in its center. 
The topological origin of such in-gap modes is also verified in Supplementary Material.
The gap closure and then the emergence of in-gap topological modes resemble the typical behavior of a topological phase transition. Yet, here it is an intriguing size-induced effect.  Further, the emergence of in-gap modes only requires exponentially weaker inter-chain coupling (i.e. smaller $\delta_{ab}/\delta_{a}$) for larger $N$, as shown in the ``phase" diagram shown in Fig.~\ref{fig:SSH}(d).


\bigskip
\noindent{\textit{Discussion.-}}
\GJ{In mathematical terms, the CSE arises when the energy eigenequation exhibits an algebraic singularity that leads to inequivalent auxiliary GBZs across the transition. The CSE heralds a whole new class of discontinuous critical phase transitions with rich anomalous scaling behavior, \CH{challenging traditional associations of criticality with scale-free behavior.}
Even a vanishingly small coupling between dissimilar skin modes can be consequential as the system size increases. This insight is much relevant to sensing and switching applications. Beyond our two-chain models, there are other scenarios that can engineer coupling between subsystems of dissimilar NHSE length scales and hence yield CSE~\cite{SuppMat}. In particular, we anticipate fruitful investigations in} {various experimentally feasible settings such as electric circuits~\cite{tobias2019observation,tobias2019reciprocal,hofmann2019chiral,ezawa2019electric}, cold atom systems~\cite{li2019topology,gou2020tunable}, photonic quantum walks~\cite{lei2019observation} and metamaterials~\cite{ananya2019observation,brandenbourger2019non}. }

\noindent{\textit{Acknowledgements.-}}
 J.G. acknowledges support from Singapore NRF Grant No. NRF-NRFI2017-04 (WBS No. R-144-000-378-281).

C.H.L. and L.L. contributed equally to this work.


%

\clearpage

\onecolumngrid
\begin{center}
\textbf{\large Supplementary Materials}\end{center}
\setcounter{equation}{0}
\setcounter{figure}{0}
\renewcommand{\theequation}{S\arabic{equation}}
\renewcommand{\thefigure}{S\arabic{figure}}
\section{I. Conditions for having discontinuous transition of GBZ solutions $E_\infty$ for the Critical Skin Effect}

\subsection{a. Two-chain models}
The discontinuous transition induced by an infinitesimal transverse coupling in thermodynamic limit, and also the crossover in a finite system, exist only when the two decoupled chains have different $\kappa$ of their OBC skin solutions. To see this, we consider a general two-chain model described by Hamiltonian
\begin{eqnarray}
h(z)=\left(\begin{matrix}
g_a(z)+V_a & t_0\\
t_0 & g_b(z)+V_b\\
\end{matrix}\right),
\end{eqnarray}
where $g_{a,b}(z)$ only contain terms with nonzero order of $z$. When decoupled, the two chains correspond to the polynomials $g_{a,b}(z)+V_{a,b}$ respectively, and possess the same $\kappa$ solutions when and only when $g_b(z)=cg_a(z)$, with $c$ a nonzero coefficient. When a nonzero transverse coupling $t_0$ is introduced, the characteristic polynomial of the two-chain system takes the form of
\begin{eqnarray}
P_c(z)&=&(g_a(z)+V_a-E)(g_b(z)+V_b-E)-r^2\nonumber\\
&=&c g_a^2(z)+\left[(V_b-E)+c(V_a-E) \right]g_a+ V_aV_b+E^2-r^2\nonumber\\
&=&(cg_a(z)-A)(g_a(z)-B),
\end{eqnarray}
where $A,B$ are two coefficients determined by $V_{a,b}$, $r=t_0$, and $E$. Therefore for two chains with the same $\kappa$ solutions, a transverse coupling $t_0$ only modifies the energy offset between them, without inducing a transition of skin solutions.

Nevertheless, the above factorization does not hold when the coupling term $r$ is $z$-dependent, corresponding to inter-chain couplings between different unit cells. Under this condition, $P_c(z)$ cannot be factorized into two sub-polynomials of $g_a(z)$ and $g_b(z)=cg_z(a)$, meaning that the skin solution is changed for the system.

\subsection{b. Dissimilar skin modes in general two-band models}
In a more general picture, the critical skin effect and the size-dependent variation may exist when different parts of the system have dissimilar skin accumulation of eigenmodes. In the two-chain model, we mainly consider regime with small inter-chain couplings, thus the two energy bands (overlapped or connected in most cases) with dissimilar skin modes are mostly given by one of the two chains respectively. To unveil the condition of having dissimilar skin modes in a general two-band system, we consider an arbitrary two-band system described by a non-Bloch Hamiltonian $H(z)=h_0(z)\mathbb{I}+\sum_{n=1,2,3}h_n(z)\sigma_n$, with $z=e^{ik}e^{-\kappa(k)}$, and $\kappa(k)$ a complex deformation of momentum $k$ describing the NHSE.
Its characteristic polynomial is given by
\begin{eqnarray}
f(z,E)=[E-h_0(z)]^2-P(z)=0
\end{eqnarray}
with $P(z)=\sum_{n=1,2,3}h_n^2(z)$.
NHSE can be described by a GBZ where the solutions of $f(z,E)=0$ satisfy $E_\alpha(z_\mu)=E_\alpha(z_\nu)$ with $|z_\mu| = |z_\nu|$ and $\alpha=\pm$ the band index, and $\kappa(k) =-\log|z|$ gives the inverse decay length.
Conventionally, NHSE is studied mostly for system with only nonzero $h_0(z)$ (i.e. a one-band model) or $P(z)$ (e.g. the non-reciprocal SSH model), where the zeros of $f(z,E)$ lead to $E_\pm=h_0(z)$ and $E_\pm^2=P(z)$ respectively. In either case, we can see that the two bands of $E_\pm$ must have the same inverse skin localization depth $\kappa(k)$, as $E_\alpha(z_\mu)=E_\alpha(z_\nu)$ must be satisfied for $\alpha=\pm$ with the same $z_{\mu,\nu}$.
To have dissimilar skin modes for the two bands, $h_0(z)$ and $P(z)$ must both be non-vanishing, and possess different skin solutions. That is, although $h_0(z_\mu)=h_0(z_\nu)$ and  $P(z_{\mu'})=h_0(z_{\nu'})$ can still be satisfied with $|z_\mu|=|z_\nu|$ and $|z_{\mu'}|=|z_{\nu'}|$, we cannot have $z_\mu=z_\mu'$ and $z_\nu=z_\nu'$ at the same time, otherwise the same $\kappa(k)$ can be obtained for the two bands.

\subsection{c. Non-monotonicity of convergence towards $E_\infty$}
\begin{figure}[H]
\includegraphics[width=1\linewidth]{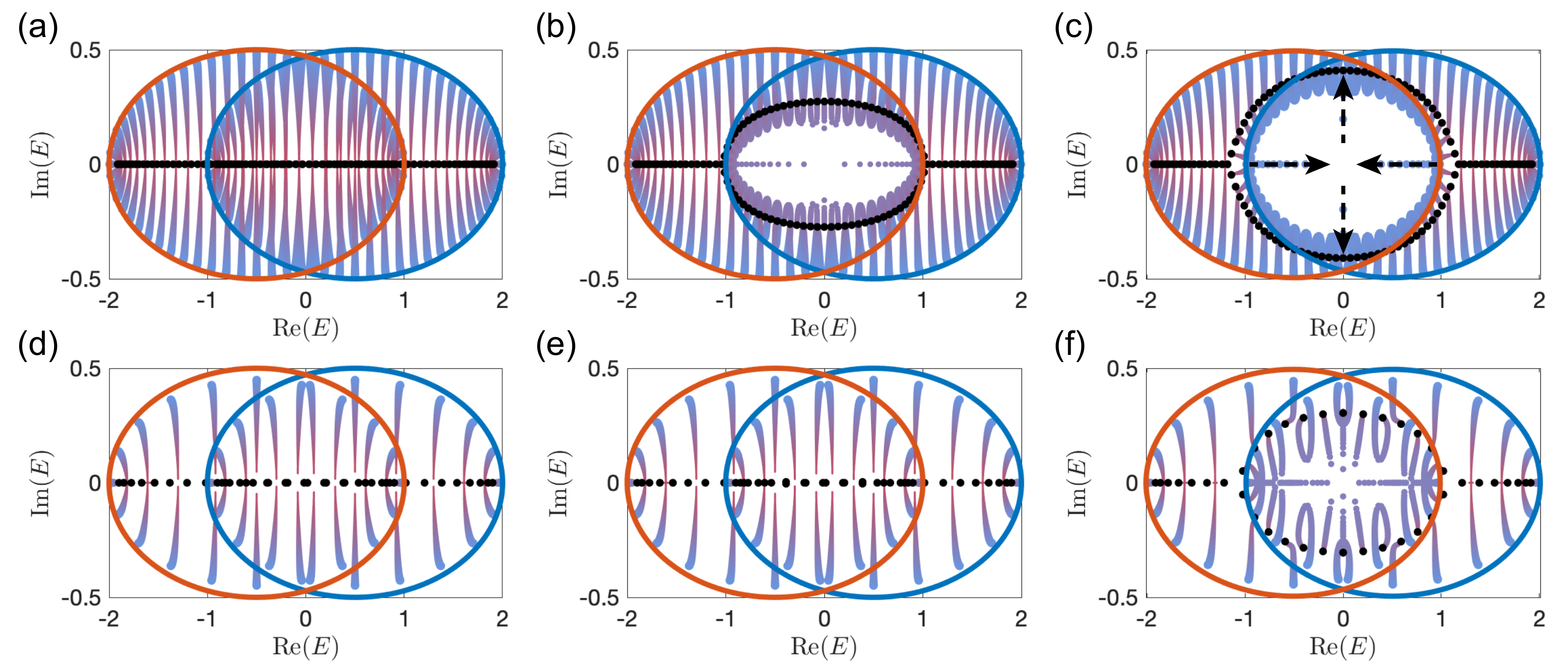}
\caption{Spectral flow of the two-chain model. (a-c) for $N=60$ unit cells, (d-f) for $N=20$ unit cells. From left to right, the inter-chain coupling is $t_0=0,2\times10^{-4},0.1$ respectively. Other parameters are $t_1=0.75$, $\delta_a=-\delta_b=0.25$, and $V=0.5$. Red and blue circles are PBC spectra obtained from the Bloch Hamiltonian, black dots are OBC spectra, and  blue-purple curves are the spectral flow from PBC to OBC. In the coupled regime of (b,c,f), two points of PBC bands on the real axis first flow toward zero energy, then rapidly separate along the imaginary axis, as shown by the arrows in Fig.~\ref{fig:flow}(c).
}
\label{fig:flow}
\end{figure}

\LH{In Fig.~\ref{fig:flow}, we illustrate the PBC-OBC spectral flow~\cite{Lee2019anatomy} of the two-chain model with different parameters, by rescaling the amplitudes of the hopping across the boundary as $t_1\pm\delta_{a,b}\rightarrow c(t_1\pm\delta_{a,b})$, and tuning $c$ from $1$ (PBC) to $0$ (OBC).
We can see that in the decoupled limit, each of the two PBC bands (red or blue) merges with itself along the real axis when approaching OBC limit [Fig.~\ref{fig:flow}(a,d,e)].
On the other hand, in the coupled regime of Fig.~\ref{fig:flow}(b,c,f), each band first flows toward the real axis, but then "turns back" and merges with the other band, forming a central-loop structure.
In this process, the PBC bands do not necessarily go monotonically closer to OBC spectrum that reflects the GBZ solutions.} \CH{A systematic study of the interplay between the switching off of boundary couplings (PBC-OBC interpolation) and subsystem coupling ($t_0$) is deferred to future work.}

\subsection{d. Reciprocal realization of the two-chain model}
\begin{figure}
\includegraphics[width=0.8\linewidth]{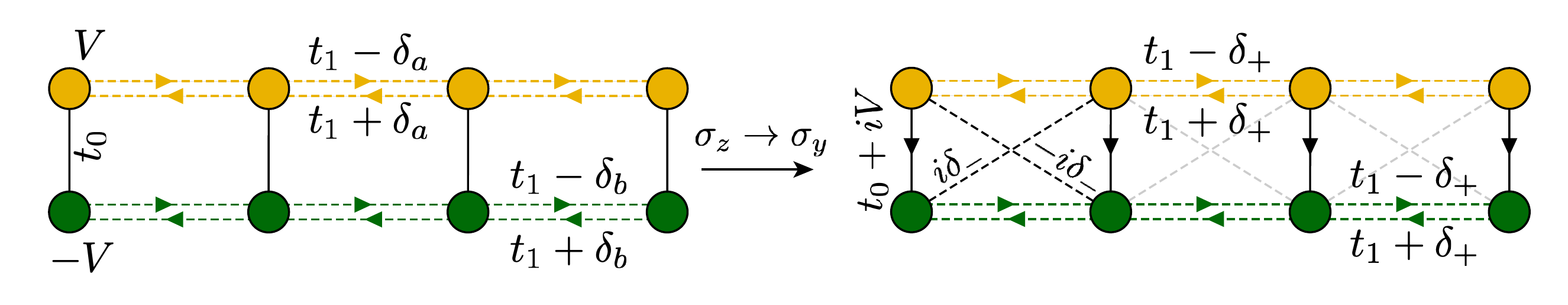}
\caption{The two-chain model with a rotation of the basis. The rotated lattice has only reciprocal hoppings when $\delta_+=V=0$.
}
\label{fig:rotated_2chain}
\end{figure}

\CH{Here, we discuss how the CSE, which requires subsystems of different NHSE decay lengths, can in fact be realized with reciprocal models that are more easily realizable in experiment.} 
In the two-chain model, the Hamiltonian can be rewritten in the form of Pauli matrices as
\begin{eqnarray}
h(z)&=&[t(z+1/z)+ \delta_+\sin k(z-1/z)]\sigma_0\nonumber\\
&&+t_0\sigma_x+[V+\delta_-(z-1/z)] \sigma_z,\label{two-chain_pauli}
\end{eqnarray}
with $\delta_\pm=(\delta_a\pm\delta_b)/2$. Here $\delta_+$ describes the equivalent part of non-Hermiticity acting on the two chains, which shall induce the same NHSE to them.
The critical behavior and transition of NHSE occurs only with nonzero $\delta_-$, which induces band-dependent NHSE along the two chains.
As shown in Fig.~\ref{fig:rotated_2chain}, $\delta_\pm$ can be divided into different couplings with a rotation of pseudospin $\sigma_z\rightarrow\sigma_y$,
and the rotated  Bloch Hamiltonian $h_r(k)$ satisfies $h_r^T(k)=h_r(-k)$ at $\delta_+=V=0$.
Under this condition, the rotated system is reciprocal, and thus provides convenience for experimental realization such as RLC circuit lattices. 

\section{II. Anomalous scaling behavior}
\subsection{a. Competition between skin localization and inter-chain coupling}
As mentioned in the main text, if two coupling chains have inverse NHSE decay lengths (non-Hermitian  localization length scales) $\kappa_a,\kappa_b$, a change of basis will bring their coupling to be effectively between a chain with no skin effect, and another with an effective skin depth $\kappa_a-\kappa_b$. Since that entails exponentially growing skin modes scaling like $e^{(\kappa_a-\kappa_b)N}$ at one end, we expect the effect of even an infinitesimally small inter-chain coupling $t_0$ to scale exponentially with $N$, and eventually change the OBC spectrum substantially.

\begin{figure}
\centering
{\includegraphics[width=.4\linewidth]{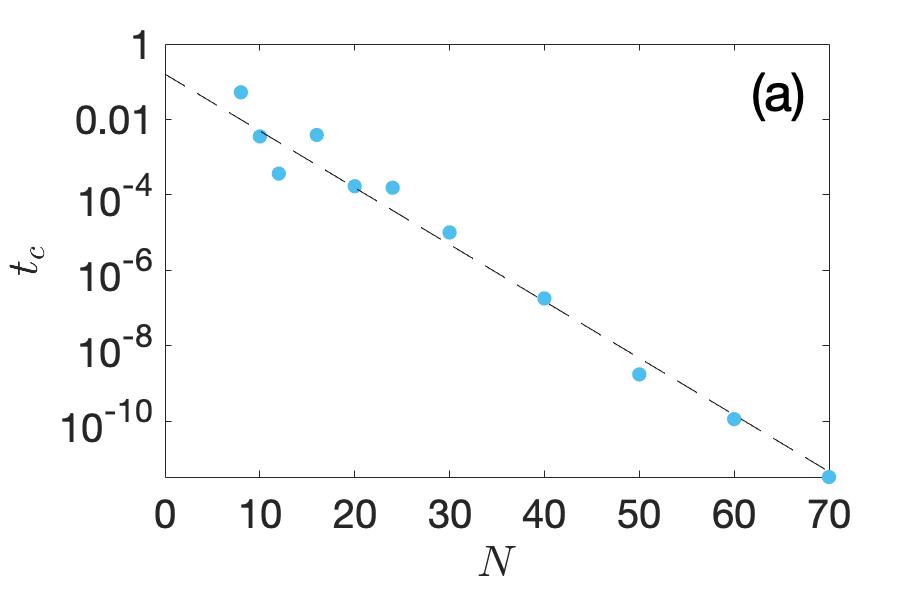}}
{\includegraphics[width=.4\linewidth]{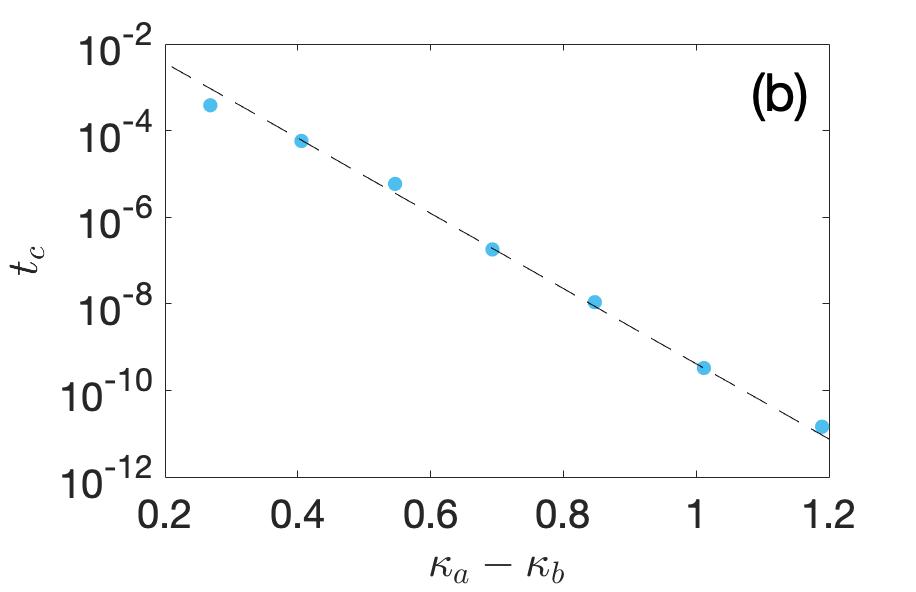}}
\caption{Inverse exponential scaling of the critical bare coupling $t_0=t_c$ required for the OBC spectrum of $H_\text{2-chain}$ to transition from real to complex, versus the system's size $N$ and effective skin depth $\kappa_a-\kappa_b$ in (a,b) respectively. The numerical data (blue) fits very well with the predicted scaling law \LH{$t_c\sim e^{-(\kappa_a-\kappa_b)N/2}$ (dashed lines)
with $\kappa_a-\kappa_b=\log 2$ in (a) and $N=40$ in (b)}.
\LH{Unless specified in the figure, the parameters are
$t_1=0.75$, $\delta_a=-\delta_b=0.25$ as in Fig.~2 of the main text.
In (b), $\kappa_a-\kappa_b$ is obtained from Eq. (\ref{eq:kappa_decoupled}) with $\delta_a=-\delta_b$ varying from $0.1$ to $0.4$.}
}
\label{fig:renorm}
\end{figure}

Consider increasing the inter-chain coupling $t_0$ in our two-chain model (Eq.~3 of main text) from zero. At sufficiently small $t_0$, we have two practically independent OBC Hatano-Nelson chains with real spectra. Their infinitesimal coupling only shifts their eigenenergies slightly along the real line. But at a critical $t_0=t_c$, the OBC spectrum is rendered complex as one or more pairs of eigenenergies coalesece and repel along in the imaginary direction. Shown in Fig.~\ref{fig:renorm}(a) is the inverse exponential scaling of the critical $t_0=t_c$ with $N$. We observe that $t_c^2e^{(\kappa_a-\kappa_b)N}\sim \mathcal{O}(1)$, in agreement with the intuitive expectation that $t_c$ should scale inverse exponentially with $N$ because the effect of $t_0$ scales exponentially with $N$. Yet, the fact that $t_c^2\sim e^{-(\kappa_a-\kappa_b)N}$ signifies that the Critical Skin Effect is fundamentally a non-perturbative effect, since it differs from $t_c\sim e^{-(\kappa_a-\kappa_b)N}$ as expected from first-order perturbation theory with left and right eigenstates that are oppositely exponentially localized spatially.

\LH{The scaling behavior of $e^{(\kappa_a-\kappa_b)N}$ also suggests that increasing $N$ has similar consequences as increasing the non-reciprocity in the system, the strength of which is reflected by the absolute value of $(\kappa_a-\kappa_b)$. Therefore it is also expected that the critical skin effect shall emerge when we enhance the non-reciprocity but fix $N$.
In Fig.~\ref{fig:renorm}(b) we show the inverse exponential scaling of the critical $t_0 = t_c$ with $\kappa_a-\kappa_b$, where the inverse NHSE decay lengths are given by
\begin{eqnarray}
e^{\kappa_{a,b}}=\sqrt{\frac{t_1+\delta_{a,b}}{t_1-\delta_{a,b}}}\label{eq:kappa_decoupled}
\end{eqnarray}
for the two decoupled chains. The scaling behavior versus $\kappa_a-\kappa_b$ further confirms that $t_c^2\sim e^{-(\kappa_a-\kappa_b)N}$.}

\LH{
In Fig.~\ref{fig:spectrum_delta-}, we illustrate another example of our two-chain model with non-Hermitian cross inter-chain coupling $\delta_{ab}$, i.e. $H_{\rm CS-SSH}$ in the main text. By increasing $\delta_-$, the non-reciprocity is strengthened along each chain, but toward opposite directions. Thus the effective inverse skin depth $\kappa_a-\kappa_b$ is enhanced, and we observe a transition of OBC spectrum from a line to a central-loop structure, accompanied with a topological transition reflected by the emergence of zero-energy degenerate edge states. This behavior is similar to the transition with enlarging $N$ as discussed in the main text.
 \begin{figure}
{\includegraphics[width=1\linewidth]{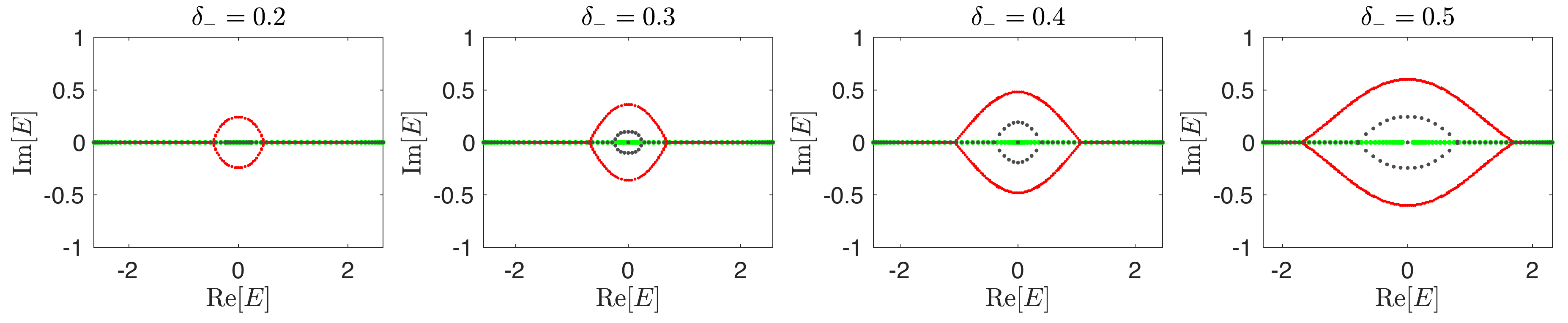}}
\caption{OBC spectra (black dots) at different $\delta_-$ for the two-chain model with non-Hermitian cross inter-chain coupling described by $H_{\rm CS-SSH}$ in the main text. Red (green) dot lines indicate the OBC skin solution in the thermodynamical limit with a weak (zero) inter-chain coupling $\delta_{ab}=0.5\times 10^{-3} (0)$. Other parameters are $N=40$, $\delta_a=-\delta_b=\delta_-$, $t_1=0.75$, and $V=1.2$.}
\label{fig:spectrum_delta-}
\end{figure}}

\subsection{b. Anomalous scaling of entanglement entropy}

The Fermionic entanglement entropy (EE) $S$ scaling behavior depends qualitatively on the nature of the phase, increasing as $~\frac1{3}\log N$ at an ordinary critical point, decreasing possibly as a negative multiple of $\log N$ at a critical exceptional point~\cite{chang2019entanglement}, and saturating at a gapped or decoupled scenario. Since $N$ itself can drive phase transitions in our case of the Critical Skin Effect, we expect the scaling of $S$ to interpolate and transition through distinct behaviors.

For free Fermions in a many-body state $|\Psi\rangle$, the (biorthogonal) EE~\cite{herviou2019entanglement,mu2019emergent} for a chosen entanglement cut can be computed via
\begin{equation}
S=-\sum_j[c_j\log c_j + (1-c_j)\log(1-c_j)],
\end{equation}
where the $c_j$'s are the eigenvalues of the 2-particle correlator $C=PQP$~\cite{peschel2003calculation,alexandradinata2011trace,lee2015free}.
Here $P$ is the projector implementing the entanglement cut and $Q=\sum_{\mu\in occ.}|\psi_\mu\rangle\langle\psi_\mu|$ is the single-body biorthogonal projector onto the set of basis states $|\psi_\mu\rangle$ occupied by the many-body state $|\Psi\rangle$. In a perfectly unentangled case, $c_j=0$ or $1$ only, giving rise to a vanishing EE. With increased entanglement, the $c_j$'s encroach closer to $1/2$, attaining the latter when the sector $j$ is fully entangled. In the biorthogonal setting, it is possible for $c_j$ to take values outside of $[0,1]$ since $|\psi_\mu\rangle$ is not the complex conjugate of $\langle \psi_\mu|$, leading to negative or even imaginary contributions to $S$~\cite{chang2019entanglement}.

\begin{figure}
\subfloat[]{\includegraphics[width=.22\linewidth]{Sodd_r=04_t=058_delta=025.pdf}}
\subfloat[]{\includegraphics[width=.22\linewidth]{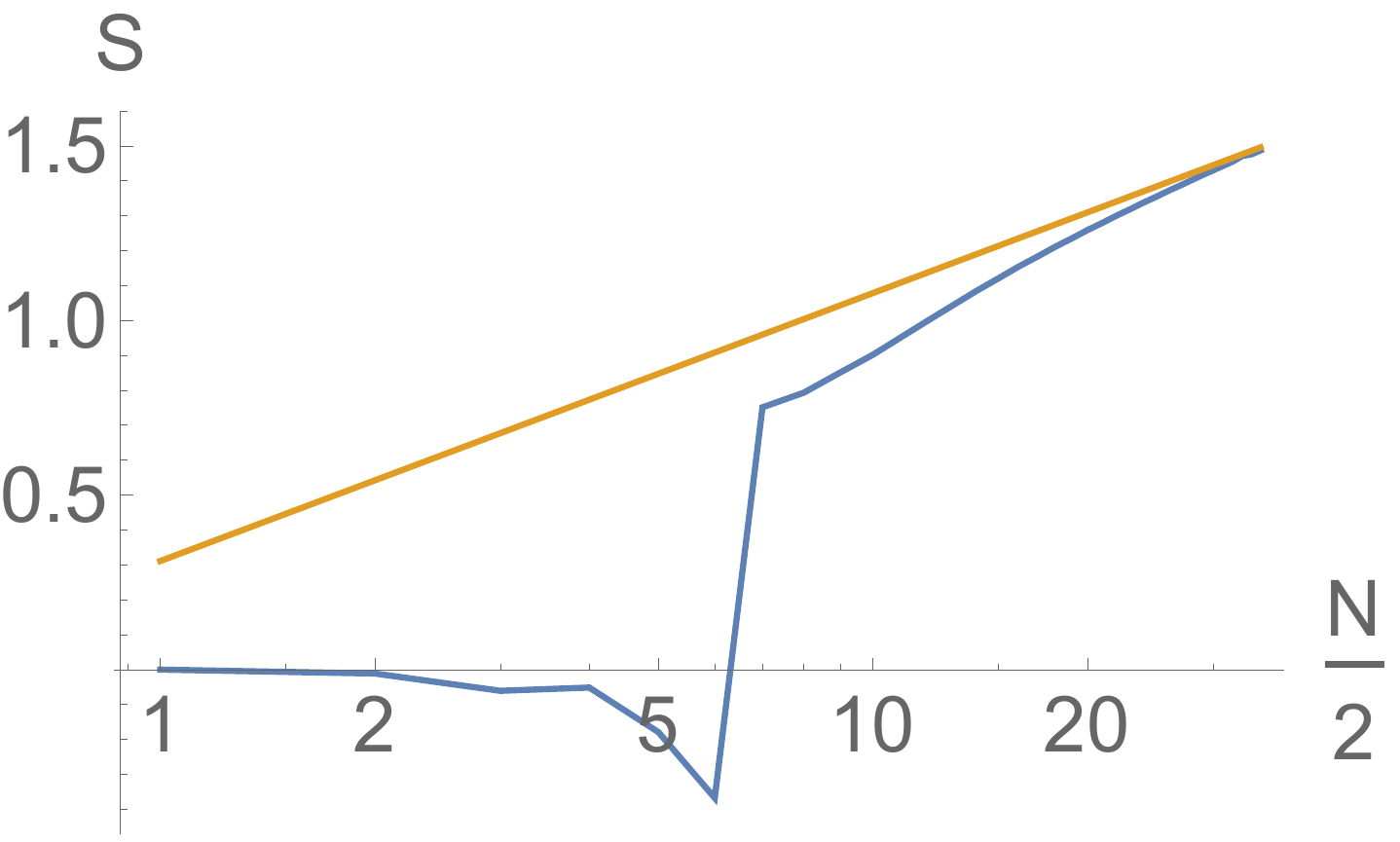}}
\subfloat[]{\includegraphics[width=.29\linewidth]{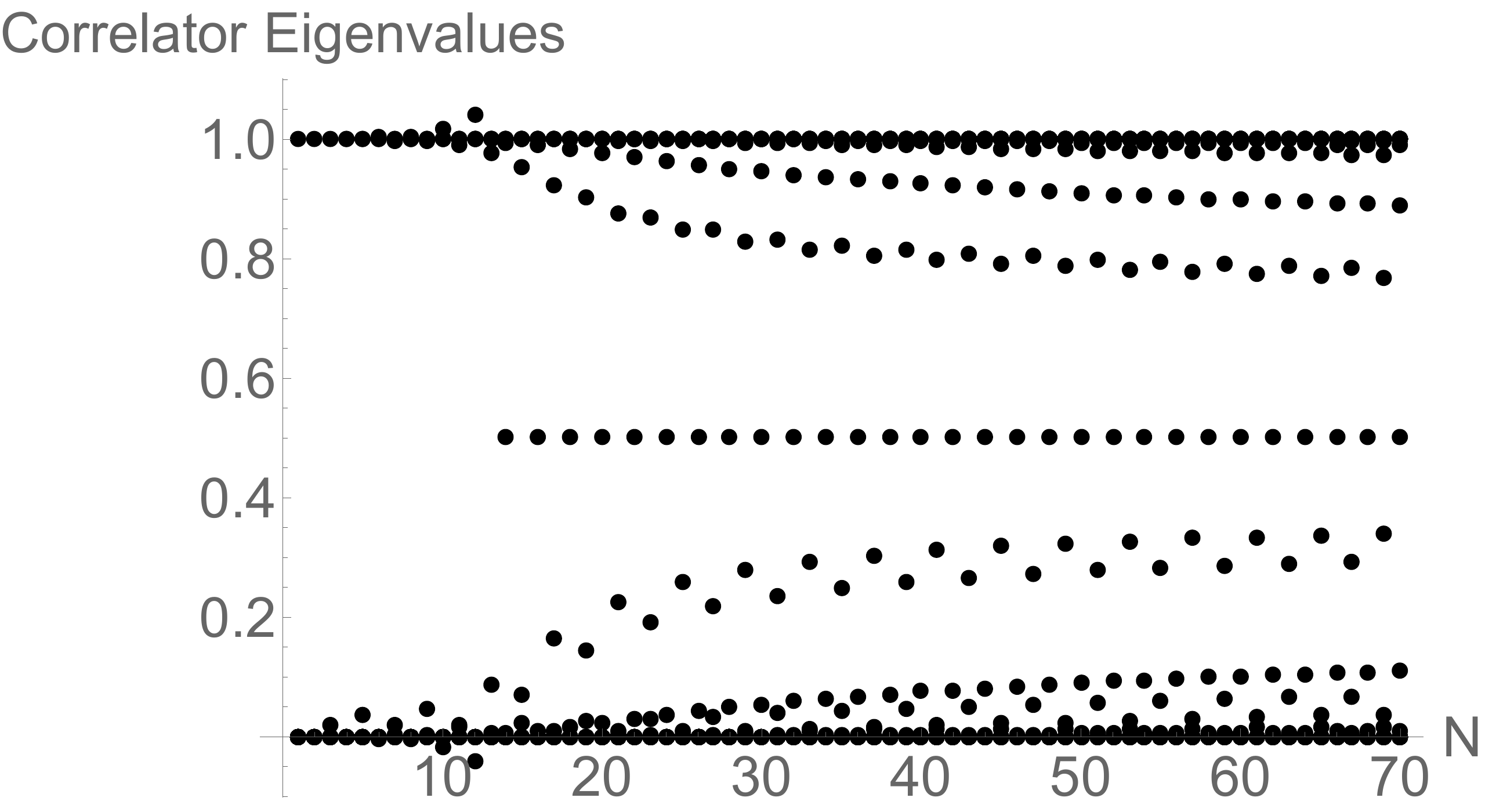}}
\subfloat[]{\includegraphics[width=.27\linewidth]{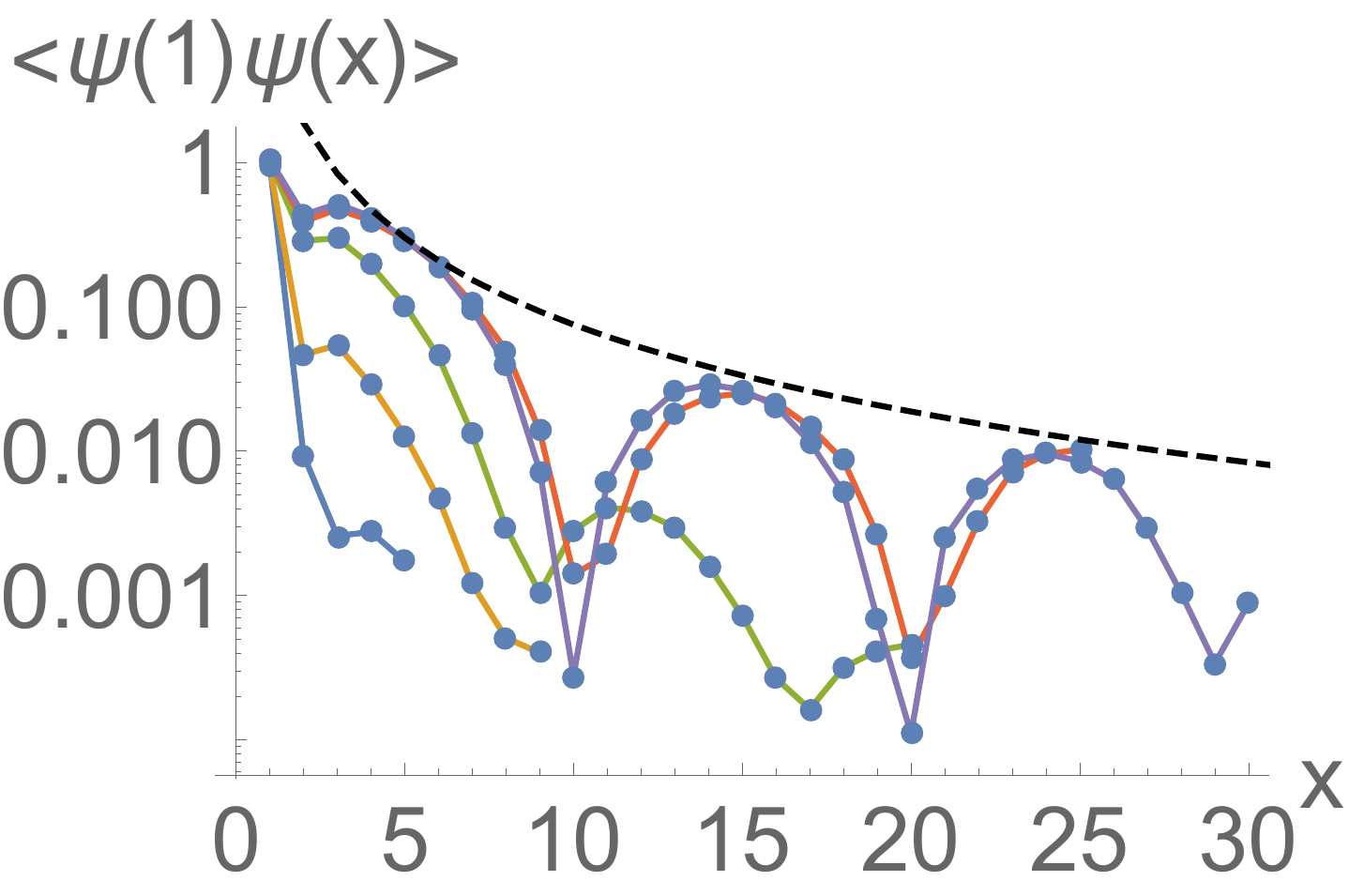}}
\caption{(a,b) Scaling of $S$ with odd/even $N$ (blue) for a half-filled OBC $H_\text{2-chain}$ with real-space cut at $\lfloor \frac{N}{2}\rfloor$ and parameters $t_1=0.58$, $V=1$, $t_0=0.4$ and $\delta_a=-\delta_b=0.25$ (same as Fig.~3a of the main text). At small $N$, $S$ is almost vanishing/is negative for odd/even system sizes. At larger $N$, both odd and even cases display a tendency towards the expected $S\sim \frac1{3}\log N$ critical behavior (yellow). (c) The corresponding correlator eigenvalues $c_j$, showing how the system transitions to critical behavior with a single $c_j=1/2$ (and other eigenvalues slowly approaching it) only beyond $N\approx 10$. Before that, the system is essentially decoupled.  (d) The corresponding two-Fermion correlation at $N=10,18,40,50,60$ (blue,brown,green,puple,red), with rapid exponential decay for small $N$ and power-law decay for large $N$ (Black dashed curve shows $N^{-1}$ decay for reference).
}
\label{fig:ent}
\end{figure}

In Fig.~\ref{fig:ent}, we observe a crossover from a decoupled regime to a critical regime when $N$ increases. $S$ also exhibits non-universal negative values for certain even $N$ [Fig.~\ref{fig:ent}(b)], \CH{a behavior resulting from $c_j\notin [0,1]$ for a few of these $N$. In real space, the two-Fermion correlator $C$ decays rapidly for small $N$, but interesting decays more slowly like $x^{-1}$ for larger $N$ when the system becomes gapless. As such, correlators generally become enhanced in larger systems where the effects of coupling become amplified by the CSE.}

\section{III. GBZ solutions $E_\infty$ for the 2-chain model}
For analytic tractability, we consider the case of Eq. 3 of the main text with $t^+_a=t_b^-=1$ and $t^-_a=t^+_b=0$ (i.e. $t_1=\delta_a=-\delta_b=0.5$), but nonzero $b$ and $V$. We obtain
\begin{equation}
H_\text{2-chain}(z)=\left(\begin{matrix}
z+V & t_0 \\
t_0 & 1/z -V
\end{matrix}\right)
\end{equation}
with characteristic polynomial given by
\begin{eqnarray}
f(z,E)&=&E^2-E(z^{-1}+z)+[(z+V)(z^{-1}-V)-t_0^2]\notag\\
&=& \frac{V-E}{z}-z(V+E)+[E^2-V^2-t_0^2+1]
\end{eqnarray}
To find the GBZ solutions $E_\infty$ for comparison with the actual OBC solutions, we solve for roots $|z_+|=|z_-|$ of $f(z,E)=0$ (with $\Sigma=E^2-V^2-t_0^2+1$):
\begin{eqnarray}
z_\pm&=&\frac{\left(\Sigma \pm\sqrt{\Sigma^2+4(V^2-E^2)}\right)}{2(V+E)}\notag\\
&=&\frac{\Sigma\pm\sqrt{(\Sigma-2)^2-4t_0^2}}{2(V+E)}
\label{zab}
\end{eqnarray}
For $|z_+|=|z_-|$ to hold, the square root quantity must differ from $\Sigma$ by a complex argument of $\pi/2$~\cite{lee2019unraveling} i.e.
\begin{equation}
\sqrt{(\Sigma-2)^2-4t_0^2}=i\GJ{\eta}\Sigma
\end{equation}
where $\eta\in \mathbb{R}$. Simplifying, we obtain $\Sigma=\frac{2}{1+\eta^2}\left(1\pm \sqrt{t_0^2+\eta^2(t_0^2-1)}\right)$ or, in terms of $E^2\rightarrow E_\infty^2$,
\begin{equation}
E_\infty^2=\frac{1-\eta^2\pm 2\sqrt{t_0^2-\eta^2+\eta^2t_0^2}}{1+\eta^2}+V^2+t_0^2
\label{Einfty}
\end{equation}
as in the main text, with $\eta$ tracing out a one-parameter continuous spectrum. The GBZ can be numerically obtained by substituting Eq.~\ref{Einfty} into the expression for $z_\pm$ in Eq.~\ref{zab} with $E=E_\infty$. From that, we obtain two momentum values $k_\pm=\text{Re}[-i\log z_\pm]$ with $\kappa(k_+)=\kappa(k_-)=-\log|z_+|=-\log|z_-|$ inverse length scales.
Note however that because of the proximity to the $t_0=0$ critical point, this value of $\kappa(k_\pm)$ is significantly different from the actual inverse OBC skin depth for a large range of finite system sizes.


\section{IV. Mapping between the SSH model and  two non-reciprocal 1D chains}

\begin{figure*}[h!]
\includegraphics[width=1\linewidth]{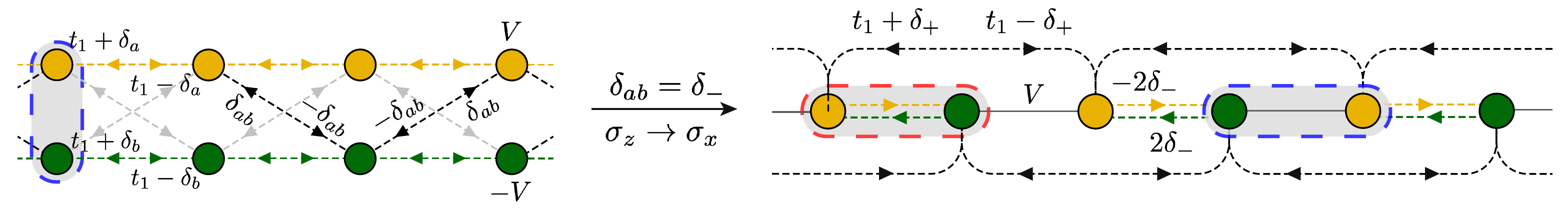}
\caption{Transforming the two-chain model with non-reciprocal cross couplings to a SSH model with non-reciprocal inter-cell couplings and second-nearest-neighbor couplings. The different parameters in the two panels are connected through $\delta_{\pm}=(\delta_a\pm\delta_b)/2$. Blue dash lines indicate a unit cell before and after the rotation, and red dash line indicates an alternative choice of unit cell with a shift of one lattice site, with which the non-reciprocal couplings of $2\delta_-$ can be further transformed into on-site gain and loss.
}
\label{fig:SSH_two-chain}
\end{figure*}

In the main text we have considered a two-chain model with both intra-chain and inter-chain couplings being non-reciprocal, described by the Hamiltonian
\begin{eqnarray}
H_\text{CS-SSH}(z)&=&[i\delta_{ab}(z+1/z)]\sigma_y+[V+\delta_-(z-1/z)]\sigma_z+[t_1(z+1/z)+\delta_+(z-1/z)]\mathbb{I}
\end{eqnarray}
with $\delta_\pm=(\delta_a\pm\delta_b)/2$.  In the parameter regime with $\delta_{ab}=\delta_-$, through a rotation of basis $\sigma_z\rightarrow\sigma_x$, this Hamiltonian becomes
\begin{eqnarray}
H_r(z)=\left(\begin{matrix}
t_1(z+1/z)+\delta_+(z-1/z) &V+2\delta_-z\\
V-2\delta_-/z & t_1(z+1/z)+\delta_+(z-1/z)
\end{matrix}\right).
\end{eqnarray}
This Hamiltonian describes a SSH model with non-reciprocal inter-cell couplings and second-nearest-neighbor couplings, as illustrated in Fig.~\ref{fig:SSH_two-chain}. In the main text we have chosen $\delta_a=-\delta_b$, so that $\delta_+=0$ and the second-nearest-neighbor couplings are Hermitian.
In this parameter regime, by redefining the unit cell as the red dashed line in Fig.~\ref{fig:SSH_two-chain} (shifting one lattice site), we can see that the rotated model is equivalent to the non-reciprocal SSH model studied in Refs.~\cite{yao2018edge,Yin2018nonHermitian} etc. with a uniform second-nearest-neighbor couplings, described by the Hamiltonian
\begin{eqnarray}
H'_{r}(z)=\left(\begin{matrix}
t_1(z+1/z) &V/z+2\delta_-\\
Vz-2\delta_- & t_1(z+1/z)
\end{matrix}\right).
\end{eqnarray}
Finally, by applying another rotation of basis $\sigma_y\rightarrow\sigma_z$, the system can be further transformed into a ladder model with non-Hermiticity being only on-site gain and loss~\cite{song2019non,li2019topology}.

Note that in the main text we have considered the case with $\delta_{ab}\ll\delta_-$. In the SSH model, this inequality corresponds to some extra longer-range couplings.
Also note that the redefinition of unit cells also corresponds to a different lattice structure where the first and last lattice sites are coupled by $V$ instead of $\pm\delta_-$. Under OBCs, these two choices of unit cells will result in different behaviors of topological edge states.

\section{V. Topological edge states in a line gap}
\begin{figure}[H]
\includegraphics[width=1\linewidth]{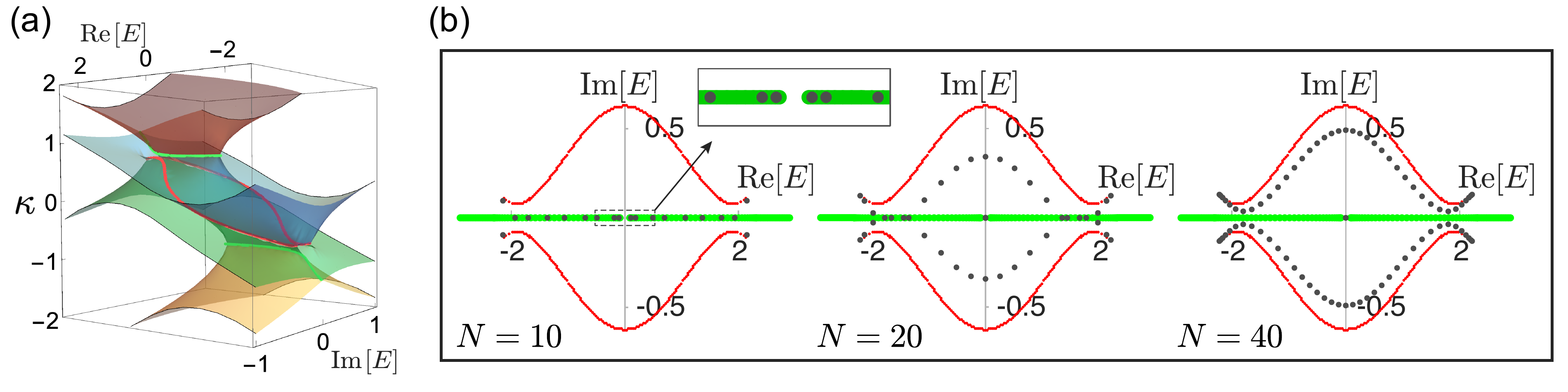}
\caption{(a) $\kappa$ solutions (red, blue, green, and yellow surfaces) of $f(z,E) = 0$ as a function of the complex energy. Parameters are $\delta_a=-\delta_b=0.5$, $t_1=0.75$, $V=1.2$, and $\delta_{ab}=0.15$. Different $\kappa$ solutions coincide along the green and red dot lines, the later one gives the OBC skin solutions of the system in the thermodynamic limit.
(b) OBC spectra (black dots) at $N = 10$, $20$ and $40$ unit cells. At small $N$, the OBC spectrum mostly lies in the real axis and is partially given by the green dot lines in (a), analogous to the skin solutions in the decoupled limit. At larger $|{\rm Re}[E]|$, however, the eigenenergies obtained different complex values and form a Y-shape spectrum, matching the OBC skin solutions of the red curves. With enlarging system's size, the spectrum continuously approaches these OBC skin solutions.
}
\label{fig:gapped_topo}
\end{figure}

Here we consider the two-chain model with cross inter-chain couplings discussed in the main text, but with a stronger inter-chain coupling strength $\delta_{ab}=0.15$.
We can see in Fig.~\ref{fig:gapped_topo} that the system has a narrow real line-gap at small $N=10$, a point-gap at $N=20$, and an imaginary line-gap at $N=40$. Degenerate zero-energy edge states emerge in the later two cases. As the two OBC bands are fully separated from each others in the last case, a Berry phase can be well-defined for each non-Bloch band to characterize the topological properties in this system.



\end{document}